\documentclass[aps,prd,twocolumn,nofootinbib]{revtex4}
\pdfoutput=1
\usepackage{epsfig}
\usepackage{amsmath}
\usepackage{graphicx}
\newcommand{\beq}{\begin{equation}}
\newcommand{\eeq}{\end{equation}}
\newcommand{\bqa}{\begin{eqnarray}}
\newcommand{\eqa}{\end{eqnarray}}

\newcommand{\bold}{\textbf}

\def\bfsigma{\mbox{\boldmath $\sigma$}}

\begin{document}

\title{Exclusive Decay of the upsilon into $h_c$, the $X(3940)$ and $X(4160)$ }
\author{Ruilin Zhu~\footnote{Email:rlzhu@sjtu.edu.cn}}
\address{$^{1}$INPAC, Shanghai Key Laboratory for Particle Physics and Cosmology,
Department of Physics and Astronomy, Shanghai Jiao Tong University, Shanghai 200240,   China\\
$^{2}$CAS Center for Excellence in Particle Physics,
Institute of High Energy Physics, Chinese Academy of Sciences, Beijing 100049, China\\
$^{3}$State Key Laboratory of Theoretical Physics, Institute of Theoretical Physics, Chinese Academy of Sciences,
Beijing 100190, China}

%\date{\today}
\begin{abstract}
In this paper, we study double charmonia production in Upsilon peaks, especially,
a S-wave charmonium $\eta_c$ and a P-wave charmonium $h_c(^1P_1)$, or a S-wave charmonium $J/\psi$ and
 the $X(3940)$ and $X(4160)$ within the nonrelativistic QCD (NRQCD)
approach which is a powerful tool to realize the factorization of double
charmonia production in electron-positron annihilation. The $J^{PC}=1^{--}$ state $\Upsilon(nS)$ can
provide an ideal laboratory for studying the properties of double-heavy quarkonium and also can
separate the perturbative and nonperturbative parts due to the large heavy quark mass compared with the typical
hadron scale $\Lambda_{QCD}$.
Explanation of the $X(3940)$ and $X(4160)$ as the $3 ^1S_0$ and $4 ^1S_0$ states, respectively,
are compatible with the observed upper limits  for the branching fractions of
$\Upsilon(1S,2S)\to J/\psi+X$, where $X=X(3940)$, $X(4160)$
 by the Belle Collaboration. The branching
fractions of $\Upsilon(1S,2S,3S)\to \eta_c+h_c(^1P_1)$ are predicted to be around $10^{-6}$,
which shall be tested in Belle-II experiments.

\begin{description}
\item[PACS numbers] 12.38.Bx,  13.25.Gv, 14.40.Pq
%\item[Keywords]
% 12.38.Bx   Perturbative calculations
% 13.25.Gv decays of jpsi,Upsilon and other quarkonia
% 14.40.Pq  heavy quarkonia
\end{description}
\end{abstract}

\maketitle

\section{Introduction}
The  $\Upsilon(nS)$ below the $B^0\bar{B}^0$ threshold, as long-lived  states with $J^{PC}=1^{--}$, have plenty of
decay modes, the measurements  which can be feasible in most current experimental techniques.
Of particular interest to
study those decay modes are the multigluon environment in the final products.
Take the $\Upsilon$ meson for instance, its dominating decay mode is the decay into three gluons,
which takes up 81.7\%~\cite{Agashe:2014kda}. Another important mode  is the radiative decay into two gluons,
which possesses 2.2\%~\cite{Agashe:2014kda}. Both of them lead to a multigluon intermediate state, and  via short-distance
hard interactions and long-distance soft effects, they furthermore transform into final measurable color-singlet
states including  conventional hadrons predicted by the constituent quark model, and  exotic states beyond
the constituent quark model such as hybrids,
multiquark states, hadron molecules, and Glueball. These $J^{PC}=1^{--}$ states with highly narrow decay width,
also including charmonia $\psi(nS)$ below the $D^0\bar{D}^0$ threshold,  have natural advantages to hunt for exotic
states since there are a great many events for these vector states that strongly couple to multigluon.  Lattice QCD
also indicates there are a large possibility for Glueball product in vector heavy quarkonium
decays~\cite{Berg:1982kp,Gui:2012gx}.  To more precisely extract the signals of the exotic states
from the conventional spectrum, however,
one need to study clearly the conventional hadrons production in those multigluon product processes.

The large bottom quark mass provides a hard scale to separate the contributions of the strong interactions
into short-distance and long-distance effects. This kind of separation is wellperformed in the nonrelativistic QCD (NRQCD)~\cite{Bodwin:1994jh}, where the amplitude
is expressed by a sum of products of non-perturbative NRQCD long-distance matrix elements (LDMEs) and  the corresponding
short-distance Wilson
coefficients. A proof of the NRQCD factorization in the exclusive double-charmonium production in an $e^+\,e^-$ annihilation,
and a quarkonium
and a light meson production in $B$ decays  was recently given in Refs.~\cite{Bodwin:2008nf,Bodwin:2010fi}.
In this paper, we employ NRQCD factorization and study the double-charmonium production in the $\Upsilon(nS)$ decays.

Double-charmonium production has attracted a large amount of attention among theorists and experimentalists
in the last ten years, especially the large discrepancy between theory~\cite{Liu:2002wq,Braaten:2002fi,Hagiwara:2003cw,Braguta:2005kr}
and data~\cite{Abe:2004ww,Aubert:2005tj} in the B factory, that is later solved  through  QCD corrections~\cite{Zhang:2005cha,He:2007te,Gong:2007db,Braguta:2008hs,Bodwin:2007ga}.
In recent years, some exclusive decay channels of bottomonium into double charmonia have been investigated
intensively, e.g., $\Upsilon\to J/\psi+\eta_c$~\cite{Jia:2007hy,Irwin:1990fn,Sang:2015owa}, $\eta_b\to J/\psi+J/\psi$~\cite{Jia:2006rx,Gong:2008ue,Braguta:2009xu,Sun:2010qx}, $\chi_{bJ}\to J/\psi+J/\psi$~\cite{Chen:2014lqa,Chen:2012ih,Sang:2011fw,Braguta:2009df}, $\Upsilon\to J/\psi+\chi_{cJ}$~\cite{Xu:2012uh}.
The branching fractions of these channels are predicted to be order from $10^{-6}$ to $10^{-8}$ in the above literatures.
In experimental aspects, the Belle Collaboration has recently measured the branching fractions for the $\Upsilon(1S,2S)$ decay
into double charmonia~\cite{Yang:2014yyy}.
Except the known charmonia, the Belle Collaboration  has also
observed the $X(3940)$~\cite{Abe:2007jna} and $X(4160)$~\cite{Abe:2007sya}  signals through the spectrum of mass recoiling against the $J/\psi$ in
the process $e^+ + e^-\to  J/\psi+X$.
Up to now, to interpret the constructions of the $X(3940)$ and $X(4160)$ states is still a challenging issue due
to lack of enough data.  In this paper, we take the $X(3940)$ as the $3 ^1S_0$ $c\bar{c}$ state and the $X(4160)$ as
the $4 ^1S_0$ $c\bar{c}$ states, and attempt to find out whether or not to explain the results observed in
the Belle experiment.
We will give the amplitudes of both $\Upsilon(nS)\to h_c(^1P_1)+\eta_c$ and $\Upsilon(nS)\to J/\psi+X$, where $X=X(3940)$, $X(4160)$.

The paper is organized as the following. In Sec.~\ref{II}, we present the NRQCD factorization calculation formulae.
The amplitudes of the channels in question are given accordingly. In Sec.~\ref{III}, the branching fractions  are
predicted and the comparisons with data are also presented. We summarize and conclude in the end section.

\section{Factorization formulae\label{II}}

\subsection{NRQCD approach}
%%%%%%%%%%%%%%%%%%%%
The NRQCD Lagrangian includes three parts, i.e. the heavy piece which describes a Schr$\ddot{o}$dinger field
for each heavy quark or antiquark, the light piece which describes ordinary QCD without  heavy quarks and heavy antiquarks, and
 the correction piece which ensures that NRQCD reproduces QCD results. It can be written as follows~\cite{Bodwin:1994jh}
%----------------------
\begin{eqnarray}
%----------------------
&&{\mathcal L}_{\rm NRQCD}=
\psi^\dagger \left( i D_t + {{\bf D}^2 \over 2m} \right) \psi+ \psi^\dagger {{\bf D}^4 \over 8m^3} \psi
%----------------------
\nonumber\\
%----------------------
&& ~~~~~+{c_F \over 2 m} \psi^\dagger \bfsigma \cdot g_s {\bf B} \psi
+ {c_D\over 8 m^2} \psi^\dagger ({\bf D}\cdot g_s {\bf E}- g_s {\bf E}\cdot {\bf D})\psi
%----------------------
\nonumber\\
%----------------------
&&~~~~~+{i c_S\over 8 m^2} \psi^\dagger \bfsigma \cdot ({\bf D}\times g_s {\bf E}- g_s {\bf E}\times {\bf D})\psi
\nonumber\\&& ~~~~~+\left(\psi \rightarrow i \sigma ^2 \chi^*, A_\mu \rightarrow - A_\mu^T\right) +
{\mathcal L}_{\rm light} \,,
%----------------------
\label{NRQCD:Lag}
%----------------------
\end{eqnarray}
%----------------------
where $\psi$ and $\chi$ denote the Pauli spinor field that annihilates a heavy quark and creates a heavy antiquark, respectively.
$D_t$ and $\bf D$ are the time and space component of the gauge-covariant derivative $D^\mu$.  $E^i=G^{0i}$ and $B^i=\frac{1}{2}\epsilon^{ijk}G^{jk}$ are the electric and magnetic color components of the gluon field strength tensor $G^{\mu\nu}$.
The replacement in the third line implies that one can obtain the corresponding
heavy antiquark bilinear sectors via the charge conjugation transformation. ${\mathcal L}_{\rm light}$ represents the Lagrangian for the light quarks and gluons.
The coefficients $c_E$, $c_F$, and $c_G$ have perturbative series in powers
of the strong coupling $\alpha_s$, which can be written as $c_i=1+{\cal O}(\alpha_s)$.

Before we write the NRQCD factorization formulae for the amplitudes of both
$\Upsilon(nS)\to h_c(^1P_1)+\eta_c$ and $\Upsilon(nS)\to J/\psi+X$, where $X=X(3940)$, $X(4160)$, we
introduce the corresponding NRQCD LDMEs.
Factorization of the inclusive annihilation decay width of heavy quarkonium can be written as
\begin{eqnarray}
\Gamma(H)=\sum_n\frac{2\mathrm{Im}f_n(\mu_\Lambda)}{m_Q^{d_n-4}}\langle H|{\cal O}_n(\mu_\Lambda)|H\rangle\,,
\end{eqnarray}
where LDMEs $\langle H|{\cal O}_n(\mu_\Lambda)|H\rangle$ involve nonperturbative effects and are well-organized by the  relative velocity $v$ between the heavy quark and heavy antiquark with the mass $m_Q$ in the heavy quarkonium $H$.
$\mathrm{Im}f_n(\mu_\Lambda)$ is the corresponding short-distance coefficient, which can be calculated order by order in perturbative theory.

According to the order counting rules, the lowest-order NRQCD operators  which contributes to the above processes we concerned are~\cite{Bodwin:1994jh}
\begin{eqnarray}
\mathcal{O}(^{1}S_{0}^{[1]})&=&\psi^{\dagger}\chi\chi^{\dagger}\psi,\\
\mathcal{O}(^{3}S_{1}^{[1]})&=&\psi^{\dagger}\bfsigma\chi\cdot\chi^{\dagger}\bfsigma\psi,\\
\mathcal{O}(^{1}P_{1}^{[1]})&=&\psi^{\dagger}(-\frac{i}{2}{\overleftrightarrow{ {\bold D}}})\chi\cdot\chi^{\dagger}(-\frac{i}{2}{\overleftrightarrow{ {\bold D}}})\psi.
\end{eqnarray}

The corresponding matrix elements of the operators sandwiched by  meson states are
\begin{eqnarray}
\langle \mathcal{O}(^{2S+1}L_{J}^{[1]})\rangle_{H} &\equiv& \langle H|\mathcal{O}(^{2S+1}L_{J}^{[1]})| H\rangle.
\end{eqnarray}

\subsection{The amplitude for $\Upsilon(nS)\to h_c+\eta_c$}

In this subsection, we calculate the amplitude for $\Upsilon(nS)\to h_c(^1P_1)+\eta_c$. Before performing the calculation,
we introduce an equivalent method~\cite{Zhang:2005cha,Jia:2007hy}, i.e. the covariant projection method, rather than the direct matching method.

The Dirac spinors for the heavy quark with momentum $p_1$ and heavy antiquark with momentum $p_2$ in quarkonium  can be written as
\begin{eqnarray}
u_Q(p_1,\lambda) &=&  \sqrt{\frac{E_1+m_Q}{2E_1}}\left(
                                           \begin{array}{ll}
                             ~~~~\xi_\lambda \\
\frac{\vec{\sigma}\cdot \overrightarrow{p_1}}{E_1+m_Q}\xi_\lambda
                                           \end{array}
                                         \right)\,,
\end{eqnarray}
\begin{eqnarray}
v_Q(p_2,\lambda) &=&  \sqrt{\frac{E_2+m_Q}{2E_2}}\left(
                                           \begin{array}{ll}
\frac{\vec{\sigma}\cdot \overrightarrow{p_2}}{E_2+m_Q}\xi_\lambda\\
                             ~~~~\xi_\lambda\end{array}
                                         \right)\,,
\end{eqnarray}
where $E_i$ are the corresponding energy of heavy quark and heavy antiquark, which satisfy $E_1=E_2\equiv E$. We introduce $q$ as half relative momentum between the heavy quark and heavy antiquark with $p_H\cdot q=0$, where $p_H=p_1+p_2$.  In the rest frame of quarkonium, we have $E=\sqrt{m_Q^2-q^2}$. $\xi_\lambda$ is the two-component Pauli spinor and $\lambda$ is the polarization parameter. Using the above formula for Dirac spinors,
it is straightforward to get the covariant form for the spin-singlet and spin-triplet combinations
of spinor bilinearities. The projections are
\begin{eqnarray}
\Pi_S(q) &=&  \sum_{\lambda_1,\lambda_2} u_Q(p_1,\lambda_1)\bar{v}_Q(p_2,\lambda_2)\langle\frac{1}{2}\lambda_1\frac{1}{2}\lambda_2|S S_z\rangle\otimes \frac{\bold{1}_c}{\sqrt{N_c}}\nonumber\\
&=&-\frac{1}{4\sqrt{2 }E(E+m_Q)}(\frac{1}{2} \,p\!\!\!\slash_{H}-q\!\!\!\slash+m_Q)\frac{p\!\!\!\slash_{H}+2E}{2E}
\nonumber\\&&\times\Gamma_S
(\frac{1}{2}\,p\!\!\!\slash_{H}+q\!\!\!\slash-m_Q)\otimes \frac{\bold{1}_c}{\sqrt{N_c}}\,,
\end{eqnarray}
where $\Gamma_{S=0}=\gamma^5$ for the spin-singlet combination, while $\Gamma_{S=1}=\varepsilon\!\!\!\slash(p_H)=\gamma^\mu \varepsilon_\mu(p_H)$ for the spin-triplet combination with the polarization
vector $\varepsilon_\mu(p_H)$, and the spin-singlet projection $\Pi_0(q)$ and the spin-triplet projection $\Pi_1(q)$
are defined accordingly.
$\bold{1}_c$ is the unit matrix in the fundamental representation of the color SU(3) group. Besides, one needs to note that the state
 $|H(\textbf{p})\rangle$ in NRQCD has the standard nonrelativistic normalization:  $\langle H(\textbf{p}^\prime)|H(\textbf{p})\rangle=(2\pi)^3\delta^3 (\textbf{p}-\textbf{p}^\prime)$, while an additional factor $2E_p$ is included in the state normalization in QCD where $\langle H(p^\prime)|H(p)\rangle=2E_p(2\pi)^3\delta^3 (\textbf{p}-\textbf{p}^\prime)$.

To obtain amplitudes of P-wave state production, one can do the  Taylor expansion of the amplitudes in powers of $q^\mu$
\begin{eqnarray}
{\cal A}(q)&=& {\cal A}(0)+\frac{\partial {\cal A}(q)}{\partial q^\mu}\mid_{q=0}q^\mu
\nonumber\\&&+\frac{1}{2!}\frac{\partial^2 {\cal A}(q)}{\partial q^\mu\partial q^\nu}\mid_{q=0}q^\mu q^\nu+\ldots.
\end{eqnarray}
\begin{figure}[th]
\begin{center}
\includegraphics[width=0.48\textwidth]{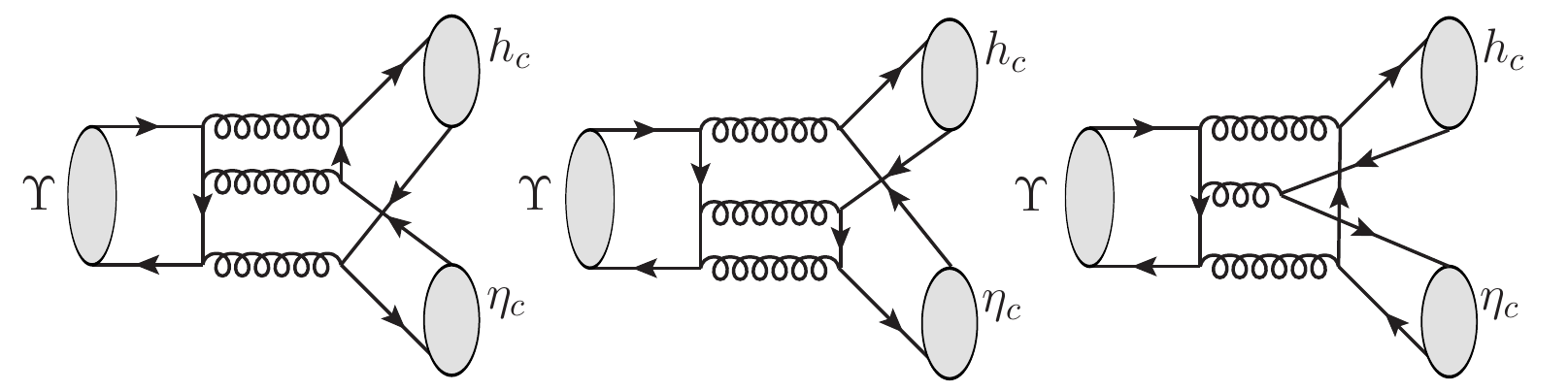}
\end{center}
    \vskip -0.7cm \caption{Typical Feynman diagrams that contribute to the process $\Upsilon(nS)\to h_c(^1P_1)+\eta_c$.}\label{Fig-feynman}
\end{figure}

The amplitude for  $\Upsilon \to h_c+\eta_c$ can be expressed as two independent terms
\begin{eqnarray}
{\cal M}(\Upsilon \to h_c+\eta_c)&=& F_1\,\varepsilon(p_\Upsilon)\cdot\varepsilon^*(p_{h_c})
\nonumber\\&&+F_0\,p_{h_c}\cdot \varepsilon(p_\Upsilon)\,p_{\Upsilon}\cdot\varepsilon^*(p_{h_c}).\nonumber\\
\end{eqnarray}
In the nonrelativistic QCD approach, the amplitude of $\Upsilon \to h_c+\eta_c$ at leading-order in $v$ can be factorized as
\begin{widetext}
\begin{eqnarray}
{\cal M}(\Upsilon \to h_c+\eta_c)&=& \sum_{L_z}N\langle 1L_z,00|1,J_z\rangle\langle 0|\chi^{\dagger}\bfsigma\psi|\Upsilon\rangle
\langle h_c|\psi^{\dagger}(-\frac{i}{2}{\overleftrightarrow{ {\bold D}}})\chi|0\rangle
\langle\eta_c|\psi^{\dagger}\chi|0\rangle
\nonumber\\&&\times\mathrm{Tr}[{\cal A}^{\mu\nu}(0)\Pi_0(0)+{\cal A}^{\nu}(0)\Pi^\mu_0(0)] \varepsilon^{*\mu} (p_{h_c},L_z)\varepsilon^\nu (p_\Upsilon),
\end{eqnarray}
where
\begin{eqnarray}
{\cal A}^{\nu}(q)&=& \frac{i\left(N_c^2-4\right) \left(N_c^2-1\right)}{128 \sqrt{2} \pi  z^2 m_b^3 N_c^{5/2} \left(\frac{p_{\eta _c}}{2}+k_2\right)^2}\int d^4 q_1\frac{\gamma ^{\alpha }.\gamma ^5.(2 z m_b+p\!\!\!\slash_{\eta_c}).\gamma ^{\beta }.(z m_b-k\!\!\!\slash_1+q\!\!\!\slash_1).\gamma ^{\delta }}{q_1^2 \left(q_1-k_1-\frac{p_{\eta _c}}{2}\right)^2 (\left(q_1-k_1\right)^2-z^2 m_b^2)}\nonumber\\
&&\times\Big[ \frac{\mathrm{Tr}[(2m_b+p\!\!\!\slash_\Upsilon).\gamma^\nu.
\gamma^\delta.(m_b+q\!\!\!\slash_1-\frac{p\!\!\!\slash_\Upsilon}{2}).\gamma^\alpha.(m_b
+q\!\!\!\slash_1-k\!\!\!\slash_1-\frac{p\!\!\!\slash_{\eta_c}}{2}+\frac{p\!\!\!\slash_\Upsilon}{2}).\gamma^\beta]}
{\left(\left(q_1-\frac{p_\Upsilon}{2}\right)^2-m_b^2\right)\left(\left(q_1-k_1-\frac{p_{\eta_c}}{2}
+\frac{p_\Upsilon}{2}\right)^2- m_b^2\right)}\nonumber\\
&&+\frac{\mathrm{Tr}[(2m_b+p\!\!\!\slash_\Upsilon).\gamma^\nu.
\gamma^\alpha.(m_b-k\!\!\!\slash_1-\frac{p\!\!\!\slash_{\eta_c}}{2}+\frac{p\!\!\!\slash_\Upsilon}{2}).\gamma^\beta.(m_b
-q\!\!\!\slash_1+\frac{p\!\!\!\slash_\Upsilon}{2}).\gamma^\delta]}
{\left(\left(q_1-\frac{p_\Upsilon}{2}\right)^2-m_b^2\right)\left(\left(k_1+\frac{p_{\eta_c}}{2}
-\frac{p_\Upsilon}{2}\right)^2- m_b^2\right)}\nonumber\\
&&+\frac{\mathrm{Tr}[(2m_b+p\!\!\!\slash_\Upsilon).\gamma^\nu.
\gamma^\beta.(m_b-k\!\!\!\slash_2-q\!\!\!\slash_1-\frac{p\!\!\!\slash_{\eta_c}}{2}+\frac{p\!\!\!\slash_\Upsilon}{2}).
\gamma^\delta.(m_b-k\!\!\!\slash_2-\frac{p\!\!\!\slash_{\eta_c}}{2}+\frac{p\!\!\!\slash_\Upsilon}{2}).\gamma^\alpha]}
{\left(\left(q_1+k_2+\frac{p_{\eta_c}}{2}
-\frac{p_\Upsilon}{2}\right)^2- m_b^2\right)\left(\left(k_2+\frac{p_{\eta_c}}{2}
-\frac{p_\Upsilon}{2}\right)^2- m_b^2\right)}\Big],\label{Amu0}
\end{eqnarray}
and
\begin{eqnarray}
&&\Pi^\mu_0(0)= \frac{\partial { \Pi_0}(q)}{\partial q^\mu}\mid_{q=0},\quad\quad\quad {\cal A}^{\mu\nu}(0)= \frac{\partial {\cal A}^{\nu}(q)}{\partial q^\mu}\mid_{q=0},
\end{eqnarray}
\end{widetext}
$N=\sqrt{2m_\Upsilon}\sqrt{2m_{h_c}}\sqrt{2m_{\eta_c}}/(2N_c)^{3/2}$ is from NRQCD operators and states normalization factor, and the vacuum-saturation approximation for LDMEs is also used: $\langle H|\mathcal{O}_n| H\rangle\simeq \langle H|\psi^{\dagger}\mathcal{K}^\prime_n\chi|0\rangle\langle 0|\chi^{\dagger}\mathcal{K}_n\psi| H\rangle$ with $\mathcal{O}_n=\psi^{\dagger}\mathcal{K}^\prime_n\chi\chi^{\dagger}\mathcal{K}_n\psi$. For convenience, we have defined $z\equiv m_c/m_b$, and the momentum $k_i$ of charm quark(antiquark) in the hadron $h_c$ can be written as $k_1=\frac{p_{h_c}}{2}+q$ and $k_2=\frac{p_{h_c}}{2}-q$.

For the $^1P_1$ charmonium state the summation over the quark spins and orbital momentum projections results in
\begin{eqnarray}
\sum_{L_z}\langle 1L_z,00|1,J_z\rangle \,\varepsilon^{*\mu} (p_{h_c},L_z)&=& \varepsilon^{*\mu} (p_{h_c},J_z).
\end{eqnarray}
The summation over the polarization of $h_c$ is
\begin{eqnarray}
\sum_{J_z=-1}^1\varepsilon^{\mu} (p_{h_c},J_z)\varepsilon^{*\nu} (p_{h_c},J_z)&=& -g^{\mu\nu}+\frac{p_{h_c}^\mu p_{h_c}^\nu}{m_{h_c}^2}.
\end{eqnarray}

Next let us give the explicit amplitude of $\Upsilon \to h_c+\eta_c$.
For pure hadronic decay, typical Feynman diagrams are depicted in Fig.~\ref{Fig-feynman}.
There are twelve diagrams contributing to the process, and another nine diagrams can be
obtained by reversing direction of each quark line one by one.  According to the power-counting rules~\cite{Bodwin:1994jh,Kramer:2001hh},
the matrix element $\mathcal{O}(^{2S+1}L_{J}^{[1,8]})_{H}$ scales as $v^{3+2L+2E+4M}$, where $S$ and $L$ are the spin and orbital angular momentum quantum number for the $Q\bar{Q}$ pair, and $E$ and $M$ are the minimum number of chromo-electric and chromo-magnetic transitions for the $Q\bar{Q}$ pair from the dominant Fock state of $H$ to the state $Q\bar{Q}(^{2S+1}L_{J}^{[1,8]})$. Consider the higher-order in $v$, the nontrivial next-to-leading order contribution for the
above process are $|b\bar{b}(^{3}S_{1}^{[1]})\rangle_{\Upsilon}\to |c\bar{c}(^{1}S_{0}^{[8]})\rangle_{h_c}+|c\bar{c}(^{1}P_{1}^{[8]})\rangle_{\eta_c}$,
$|b\bar{b}(^{3}P_{J=0,1,2}^{[8]})\rangle_{\Upsilon}\to |c\bar{c}(^{1}S_{0}^{[8]})\rangle_{h_c}+|c\bar{c}(^{1}S_{0}^{[1]})\rangle_{\eta_c}$, and $|b\bar{b}(^{3}P_{J=0,1,2}^{[8]})\rangle_{\Upsilon}\to |c\bar{c}(^{1}P_{1}^{[1]})\rangle_{h_c}+|c\bar{c}(^{1}P_{1}^{[8]})\rangle_{\eta_c}$, where the
first one has a relative suppression $v^4_{c\bar{c}} $, while the last two have
a relative suppression $v^2_{b\bar{b}} v^2_{c\bar{c}}$ compared to the color-singlet contribution.

The diagrams in Fig.~\ref{Fig-feynman} together with other nine topological diagrams give  the corresponding coefficients
\begin{widetext}
\begin{eqnarray}
F_0&=& -\frac{\pi  C_F  (2 N_c C_F-3 )\alpha_s^3}{3 z^3 (1-4 z^2)^2  (3 z^2-1 ) m_b^{13/2} N_c^3}\langle 0|\chi^{\dagger}\bfsigma\psi|\Upsilon\rangle\langle h_c|\psi^{\dagger}(-\frac{i}{2}{\overleftrightarrow{ {\bold D}}})\chi|0\rangle\langle\eta_c|\psi^{\dagger}\chi|0\rangle\nonumber\\
&&\times\{1920 z^{10}-4256 z^8+432 z^6+324 z^4-37 z^2+3+(-1792 z^8+640 z^6-20 z^4\nonumber\\
&&-13 z^2+3)b_1+12 z^2 (80 z^8-44 z^6-22 z^4+13 z^2-1 )b_2+2 z^2  (3 z^2-1 ) (152 z^4\nonumber\\
&&+38 z^2-1)b_3+4 z^2  (340 z^6-147 z^4+30 z^2-7 )b_4+(3 z^2 (-320 z^8+16 z^6\nonumber\\
&&+96 z^4-58 z^2+17)-3)b_5+6 z^2  (3 z^2-1 ) (16 z^6+20 z^4+10 z^2-1 )c_1\nonumber\\
&&-12 z^2 (3 z^2-1 ) (8 z^6-6 z^4+6 z^2+1 )c_2\},
\end{eqnarray}
\begin{eqnarray}
F_1&=& \frac{2 \pi  C_F \left(2 N_c C_F-3\right)\alpha_s^3}{3 z^3 \left(4 z^2-1\right) m_b^{9/2} N_c^3}\langle 0|\chi^{\dagger}\bfsigma\psi|\Upsilon\rangle\langle h_c|\psi^{\dagger}(-\frac{i}{2}{\overleftrightarrow{ {\bold D}}})\chi|0\rangle \langle\eta_c|\psi^{\dagger}\chi|0\rangle\nonumber\\
&&\times\{112 z^4+8 z^2+3+(16 z^4+8 z^2+3)b_1+12 z^2  (4 z^2-1 )b_2\nonumber\\&&-2 z^2  (20 z^2+1 )b_3+24 z^2  (z^2-1 )b_4+(-48 z^4+30 z^2-3)b_5\nonumber\\&&
-6 z^2  (8 z^2-1 )c_1-12 z^2  (5 z^2-2 )c_2\},
\end{eqnarray}
\end{widetext}
where the one-loop master integrals $b_i$ and $c_i$ are presented in the appendix.
\subsection{The amplitude for $\Upsilon(nS)\to J/\psi+X(M)$}
Considering the Belle Collaboration has observed the channels of $\Upsilon(nS)\to J/\psi+X$ where
$X$ can be either one of the $X(3940)$ and $X(4160)$, we go to study these channels.
The $X(3940)$~\cite{Abe:2007jna} and $X(4160)$~\cite{Abe:2007sya} are first observed by
the Belle Collaboration in the spectrum of mass recoiling against the $J/\psi$ in
the process $e^+ + e^-\to  J/\psi+X$. Up to now, we have no more internal
information except the fact that both of them shall have a charm quark-antiquark pair because of their hadronic decays into $D\bar{D}$ or $D^*\bar{D}$. There are some different schemes to interpolate the nature of the $X(3940)$ and $X(4160)$, e.g. a hybrid charmonium~\cite{Petrov:2005tp}, a molecular charmonium~\cite{Fernandez:2014pla}, and a pure charmonium state~\cite{Braguta:2006py,Chao:2007it,Yang:2009fj,Sreethawong:2014jra}.

In this paper, we take in a pure charmonium explanation for these states, and  adopt the $3 ^1S_0$ and $4 ^1S_0$ states to interpolate the $X(3940)$ and $X(4160)$, respectively. Then we just  calculate the process of $\Upsilon(nS)\to J/\psi+c\bar{c}(3 ^1S_0,\,4^1S_0)$.
The pioneer works of $\Upsilon(nS)\to J/\psi+\eta_c$ have been performed in Ref.~\cite{Jia:2007hy,Irwin:1990fn}. We have also calculated
the amplitudes, which are in agreement with Ref.~\cite{Jia:2007hy}.

Following the same procedure, the amplitude of $\Upsilon \to J/\psi+X(3940)$ at leading-order in $v$ can be factorized as
\begin{eqnarray}
&&{\cal M}(\Upsilon \to J/\psi+X(3940))\nonumber\\&=& N^\prime\langle 0|\chi^{\dagger}\bfsigma\psi|\Upsilon\rangle
\rangle\langle J/\psi|\psi^{\dagger}\bfsigma\chi|0\rangle\nonumber\\&&\times\langle X(3940)|\psi^{\dagger}\chi|0\rangle \mathrm{Tr}[{\cal A}^{\nu}(0)\Pi^\mu_1(0)_{J/\psi}]
\nonumber\\&& \times\varepsilon^{*\mu} (p_{J/\psi})\varepsilon^\nu (p_\Upsilon),
\end{eqnarray}
where $N^\prime=\sqrt{3}\sqrt{2m_\Upsilon}\sqrt{2m_{J/\psi}}\sqrt{2m_{X}}/(2N_c)^{3/2}$ and ${\cal A}^{\nu}(0)$ are identical to
Eq.~(\ref{Amu0}) at $q=0$.

The amplitude for  $\Upsilon \to J/\psi+X(3940)$ can be written as
\begin{eqnarray}
&&{\cal M}(\Upsilon \to J/\psi+X(3940))\nonumber\\
&=& iF_A\epsilon_{\mu\nu\alpha\beta}\,\varepsilon^\mu (p_\Upsilon) \varepsilon^{*\nu}(p_{J/\psi})\,p_{\Upsilon}^\alpha p_{h_c}^\beta.
\end{eqnarray}

The diagrams contributing to $\Upsilon \to J/\psi+X(3940)$ are analogous to $\Upsilon \to h_c+\eta_c$, which can be got by replacing $h_c$ to $J/\psi$ and $\eta_c$ to $X(3940)$ in Fig.~\ref{Fig-feynman}. We obtain the coefficient
\begin{widetext}
\begin{eqnarray}
F_A&=& \frac{\pi  C_F \left(2 N_c C_F-3\right)\alpha_s^3}{z^2 \left(4 z^2-1\right) m_b^{11/2} N_c^3}\langle 0|\chi^{\dagger}\bfsigma\psi|\Upsilon\rangle\langle J/\psi|\psi^{\dagger}\bfsigma\chi|0\rangle\langle X(3940)|\psi^{\dagger}\chi|0\rangle\nonumber\\
&&\times\{-16 z^4+12 z^2-1+(4 z^2-1)b_1-4 z^2 \left(2 z^2-1\right)b_2-2 z^2b_3\nonumber\\
&&\times-4 z^2b_4+(8 z^4-2 z^2+1)b_5+2 z^2 \left(2 z^2-1\right)c_1+8 z^4c_2\}.
\end{eqnarray}
\end{widetext}
Because we treat the $X(3940)$ and $X(4160)$ as the $3 ^1S_0$ and $4 ^1S_0$ states separately, the coefficient for $\Upsilon \to J/\psi+X(4160)$ can  also be obtained by just replacing the LDME $\langle 0|\chi^{\dagger}\psi|X(3940)\rangle$ to $\langle 0|\chi^{\dagger}\psi|X(4160)\rangle$.
\section{Phenomenological discussions\label{III}}

Experimental data by the Belle Collaboration indicate that the cross section of
$e^++e^-\to J/\psi+X$  where $X=X(3940)$, $X(4160)$ is not trivial~\cite{Abe:2007jna,Abe:2007sya}. One has to resort to certain schemes to describe the nature of the  $X(3940)$ and $X(4160)$, and different treatments determine
distinctive results. A pure charmonium explanation for both the  $X(3940)$ and $X(4160)$ are popular
in current literatures~\cite{Chao:2007it,Yang:2009fj,Sreethawong:2014jra}. Of course,  the measurement
with more data will tell us the nature inside  the  $X(3940)$ and $X(4160)$. Except for the  $X(3940)$
and $X(4160)$, measurements of double charmonia production with a S-wave charmonium
and a P-wave charmonium are also feasible in current Belle experiment~\cite{Yang:2014yyy}.

In the above section, we have given  the NRQCD factorization formulae and calculated the corresponding short-distance coefficients. Next we input the value of
LDMEs and other fundamental parameters. In this paper, the $X(3940)$ is treated as the $\eta_c(3 S)$,
and the $X(4160)$ is treated as the $\eta_c(4 S)$.  For the LDMEs of $\eta_c(nS)$, we can get relation
the relationship of $\psi(nS)$  using the heavy quark spin symmetry. In the leading-order of heavy quark
relative velocity $v$, they are identical. So the LDMEs for the  $X(3940)$ and $X(4160)$ can be obtained
correspondingly. For the LDMEs of $\psi(nS)$, we can obtain the value from the measured electric widths~\cite{Bodwin:1994jh,Qiao:2012hp,Qiao:2014pfa},
i.e.
\begin{eqnarray}
\langle \mathcal{O}(^{3}S_{1}^{[1]})\rangle_{J/\psi}=\frac{N_c m^2_{J/\psi }}{8\pi\alpha^2e_{c}^2}\frac{
\Gamma (J/\psi\rightarrow e^+ e^-)}{(1-4\alpha_s C_F/\pi)}.\label{jpsildms}
\end{eqnarray}
Using the above expression, the corresponding values of both LO and NLO for the LDMEs are given in Tab.~\ref{tab:LDMEs}.

The parameters we adopted are~\cite{Agashe:2014kda}
\begin{center}
  $m_Y=9.4603${GeV},~$m_{Y(2S)}=10.023${GeV},~\\$m_{Y(3S)}=10.355${GeV},~$m_{Y(4S)}=10.579${GeV},\\
  $m_{Y(5S)}=10.876${GeV},~$m_{\eta_c}=2.984${GeV},~\\$m_{\eta_c(2S)}=3.639${GeV},~$m_{J/\psi}=3.097${GeV},\\
  $m_{\psi(2S)}=3.686${GeV},~$m_{h_c}=3.525${GeV},~\\$m_{\psi(4040)}=4.039${GeV},~$m_{\psi(4415)}=4.421${GeV},\\
  $\Gamma(\Upsilon)=54.02${keV},~$\Gamma(\Upsilon(2S))=31.98${keV},~\\$\Gamma(\Upsilon(3S))=20.32${keV},
  ~$\Gamma_{ee}(\Upsilon)=1.285${keV},\\
$\Gamma_{ee}(\Upsilon(2S))=0.612${keV},~$\Gamma_{ee}(\Upsilon(3S))=0.443${keV},\\$\Gamma_{ee}(\Upsilon(4S))=0.272${keV},
$\Gamma_{ee}(\Upsilon(5S))=0.31${keV},\\~$\Gamma_{ee}(J/\psi)=5.55${keV},~$\Gamma_{ee}(\psi(2S))=2.36${keV},\\
$\Gamma_{ee}(\psi(4040))=0.86${keV},~$\Gamma_{ee}(\psi(4415))=0.58${keV}.
\end{center}
For the mass of the $X(3940)$ and $X(4160)$, we have $m_{X(3940)}=3.942${GeV} and  $m_{X(4160)}=4.156${GeV}.
The heavy quark mass is adopted as $m_c=1.5${GeV} and $m_b=4.8${GeV}~\cite{Qiao:2012vt,Qiao:2011zc}. The strong
coupling constant is set at the Z-boson point with  $\alpha_s(m_Z)=0.1185$ where $m_Z=91.1876${GeV}~\cite{Agashe:2014kda}, so we can run the coupling to other points. For example, $\alpha_s(4.8\mathrm{GeV})=0.2178$ at two-loop evolution with active flavor $n_f=5$.

\begin{table}[thb]
\caption{\label{tab:LDMEs} The value of LDMEs for $\psi(nS)$ and $\Upsilon(nS)$ extracted from the electric widths at LO and NLO, in comparison with the results from B-T potential model by Buchm$\ddot{u}$ller and Tye~\cite{Buchmuller:1980su,Eichten:1995ch}. The strong coupling constant is evaluated at the scale $2m_Q$,  and varying the scale from $m_Q/2$ to the meson mass we can get the corresponding uncertainties.}
\begin{center}

\begin{tabular}{cccc}
\hline\hline
 $ (\mathrm{GeV})^3$& LO&NLO&  B-T model\\
\hline
$\langle \mathcal{O}(^{3}S_{1}^{[1]})\rangle_{J/\psi}$ & 0.2341 & $0.4107^{+0.1378}_{-0.0034}$ & 0.3867  \\
$\langle \mathcal{O}(^{3}S_{1}^{[1]})\rangle_{\psi(2S)}$ & 0.1411 & $0.2475^{+0.0831}_{-0.0198}$ & 0.2526  \\
$\langle \mathcal{O}(^{3}S_{1}^{[1]})\rangle_{\psi(3S)}$ & 0.0617 & $0.1083^{+0.0363}_{-0.0715}$ & 0.2172  \\
$\langle \mathcal{O}(^{3}S_{1}^{[1]})\rangle_{\psi(4S)}$ & 0.0499 & $0.0875^{+0.0294}_{-0.0719}$ & --  \\
$\langle \mathcal{O}(^{3}S_{1}^{[1]})\rangle_{\Upsilon}$ & 2.025 & $2.925^{+0.288}_{-0.0}$ & 3.093  \\
$\langle \mathcal{O}(^{3}S_{1}^{[1]})\rangle_{\Upsilon(2S)}$ & 1.082 & $1.563^{+0.154}_{-0.007}$ & 1.544  \\
$\langle \mathcal{O}(^{3}S_{1}^{[1]})\rangle_{\Upsilon(3S)}$ & 0.8361 & $1.208^{+0.119}_{-0.009}$ & 1.181  \\
$\langle \mathcal{O}(^{3}S_{1}^{[1]})\rangle_{\Upsilon(4S)}$ & 0.6340 & $0.9158^{+0.0902}_{-0.0117}$ & 1.025  \\
$\langle \mathcal{O}(^{3}S_{1}^{[1]})\rangle_{\Upsilon(5S)}$ & 0.6413 & $0.9262^{+0.0912}_{-0.0118}$ & 0.9339  \\
\hline\hline
\end{tabular}
\end{center}
\end{table}

For the LDMEs of $\eta_c(nS)$, we can obtain the following approximation using the heavy quark spin symmetry
\begin{eqnarray}
\langle \mathcal{O}(^{1}S_{0}^{[1]})\rangle_{\eta_c(nS)}\simeq\langle \mathcal{O}(^{3}S_{1}^{[1]})\rangle_{\psi(nS)}.
\end{eqnarray}
To take $\psi(4040)$ and $\psi(4415)$ as $\psi(4S)$ and $\psi(5S)$ respectively~\cite{Ebert:2011jc}, we have
\begin{eqnarray}
\langle \mathcal{O}(^{1}S_{0}^{[1]})\rangle_{\eta_c(3S)}&=&\langle \mathcal{O}(^{3}S_{1}^{[1]})\rangle_{\psi(4040)}
\nonumber\\&=&0.1083^{+0.0363}_{-0.0715} (\mathrm{GeV})^3,\nonumber\\
\langle \mathcal{O}(^{1}S_{0}^{[1]})\rangle_{\eta_c(4S)}&=&\langle \mathcal{O}(^{3}S_{1}^{[1]})\rangle_{\psi(4415)}
\nonumber\\&=&0.0875^{+0.0294}_{-0.0719} (\mathrm{GeV})^3.\nonumber\\
\end{eqnarray}

\begin{table}[t]
\caption{\label{tab:data} Branching fractions ($10^{-6}$) for $\Upsilon(nS)$ hadronic decays into double charmonia, where the first column uncertainty comes from  the scale running from $2m_c$ to $2m_b$ and the second column uncertainty is from the heavy quark mass with $m_c=1.5\pm0.1${GeV} and $m_b=4.8\pm0.1${GeV}. For comparison,  the results ${\cal{B}}_{\mathrm{Jia}}$ are from Ref.~\cite{Jia:2007hy}, and  the data are from the Belle experiment~\cite{Yang:2014yyy}.}
\begin{center}
\begin{tabular}{cccc}
\hline\hline
 Channels & ${\cal{B}}_{\mathrm{NRQCD}}$ & ${\cal{B}}_{\mathrm{Jia}}$ & ${\cal{B}}_{\mathrm{Exp}}$ \\
\hline
$\Upsilon \to J/\psi+\eta_c$ & $3.92^{+5.77+0.59}_{-2.62-0.53}$ & $3.9^{+5.6}_{-2.3}$ & $<$2.2  \\
$\Upsilon \to J/\psi+\eta_c(2S)$ & $1.77^{+2.61+0.27}_{-1.18-0.24}$ &$2.0^{+3.4}_{-1.4}$& $<$2.2   \\
$\Upsilon \to J/\psi+X(3940)$ & $0.68^{+1.00+0.10}_{-0.45-0.09}$ & --& $<$5.4   \\
$\Upsilon \to J/\psi+X(4160)$ & $0.49^{+0.74+0.07}_{-0.33-0.07}$ &--& $<$5.4   \\
$\Upsilon \to \eta_c+h_c$ &$1.33^{+1.96+0.34}_{-0.89-0.26}$ &--& --  \\
$\Upsilon \to \psi(2S)+\eta_c$ &  $2.18^{+3.21+0.33}_{-1.46-0.29}$ &$1.7^{+2.4}_{-1.0}$& $<$3.6   \\
$\Upsilon \to J/\psi+X(4160)$ & $0.43^{+0.64+0.06}_{-0.29-0.06}$ &--&$<$2.0   \\
$\Upsilon(2S) \to \eta_c+h_c$ & $1.12^{+1.65+0.29}_{-0.75-0.22}$ &--&--   \\
$\Upsilon(3S) \to J/\psi+\eta_c$ &$3.80^{+5.59+0.57}_{-2.54-0.51}$ &--& --  \\
$\Upsilon(3S) \to\psi(2S)+\eta_c(2S)$ &  $0.97^{+1.42+0.14}_{-0.64-0.13}$ &$0.8^{+1.4}_{-0.6}$& $<$3.2   \\
$\Upsilon \to \psi(2S)+X(3940)$ & $0.36^{+0.54+0.05}_{-0.24-0.05}$&-- & $<$2.9  \\
$\Upsilon \to \psi(2S)+X(4160)$ & $0.26^{+0.39+0.04}_{-0.18-0.04}$ &--& $<$2.9 \\
$\Upsilon(2S) \to J/\psi+\eta_c$ & $3.27^{+4.82+0.49}_{-2.19-0.44}$ &$2.6^{+3.7}_{-1.6}$& $<$5.4  \\
$\Upsilon(2S) \to J/\psi+\eta_c(2S)$ & $1.39^{+2.05+0.21}_{-0.93-0.19}$&$1.3^{+2.1}_{-0.9}$ & $<$2.5   \\
$\Upsilon(2S) \to J/\psi+X(3940)$ & $0.58^{+0.86+0.09}_{-0.39-0.08}$ &--& $<$2.0  \\
$\Upsilon(2S) \to J/\psi+\eta_c(2S)$ & $1.64^{+2.41+0.25}_{-1.09-0.22}$&-- & --   \\
$\Upsilon(3S) \to J/\psi+X(3940)$ & $0.68^{+1.00+0.10}_{-0.46-0.09}$&--& --   \\
$\Upsilon(3S) \to J/\psi+X(4160)$ & $0.51^{+0.75+0.08}_{-0.34-0.07}$&-- & --   \\
$\Upsilon(3S) \to \eta_c+h_c$ & $1.31^{+1.92+0.33}_{-0.87-0.25}$ &--& --  \\
\hline\hline
\end{tabular}
\end{center}
\end{table}

For the LDMEs of $h_c$, we adopt the result of B-T potential model in Refs.~\cite{Buchmuller:1980su,Eichten:1995ch}
\begin{eqnarray}
\langle \mathcal{O}(^{1}P_{1}^{[1]})\rangle_{h_c}=0.1074(\mathrm{GeV})^5.
\end{eqnarray}

The decay width of $\Upsilon \to h_c+\eta_c$ can be written as:
\begin{eqnarray}
\Gamma(\Upsilon \to h_c+\eta_c)&=&\frac{|\textbf{p}|}{8\pi
m_{\Upsilon}^2}|\overline{{\cal M}}(\Upsilon \to h_c+\eta_c)|^2,
\end{eqnarray}
where
\begin{eqnarray}
|\textbf{p}|&=&\frac{\sqrt{\left(m_{\Upsilon }^2-\left(m_{\eta _c}-m_{h_c}\right){}^2\right) \left(m_{\Upsilon }^2-\left(m_{h_c}+m_{\eta _c}\right){}^2\right)}}{2 m_{\Upsilon }},\nonumber
\end{eqnarray}
is the momentum modulus of final charmonia in the $\Upsilon$ meson rest frame, and
${\cal \overline{M}}^2={\cal M}^2/3$ is the amplitude with average polarization.

In the end, we give  the branching fractions of the processes we concerned,
which are presented in Tab.~\ref{tab:data}. The Belle data
are also listed for comparison. The results we presented can explain the
experimental upper limits for the processes in question.
And we also predict the branching fractions of
$\Upsilon(nS) \to h_c+\eta_c$ and $\Upsilon(3S) \to J/\psi+X(M)$,
which have the potential to be observed in current experiment.
Note that in  Ref.~\cite{Jia:2007hy}, Jia has investigated the process $\Upsilon \to J/\psi+\eta_c$ including both the three-gluon mode and the electromagnetic radiative mode contributions. However, the electromagnetic radiative mode contributions are small as the author has pointed out.

\section{Conclusion}
In this paper, we have calculated the branching fractions of $\Upsilon(nS) \to h_c+\eta_c$ and
$\Upsilon(nS) \to J/\psi+X$ with $X=X(3940)$ or $X=X(4160)$ using the NRQCD factorization approach.
For the Upsilon peaks below the $B^0\bar{B}^0$ threshold, the branching fractions of the processes
in question are around $10^{-6}-10^{-7}$; thus, these channels have the potential to be measured
in the  Belle-II experiment, where the events of $\Upsilon$, $\Upsilon(2S)$, and $\Upsilon(3S)$ will go to $1.8\times 10^{11}$,
$7.0\times 10^{10}$, and $3.7\times 10^{10}$, respectively~\cite{Shen2015}.  At Upsilon peaks higher than the $B^0\bar{B}^0$ threshold,
the branching fractions will be suppressed by a factor around $10^3$, because the decay width of the
corresponding $\Upsilon(nS)$ meson increases greatly up to dozens of {MeV}. The Upsilon family,
also existing in the $J/\psi$ family, provides a natural laboratory for multigluon products and
indicates the potential existing space for the exotic states beyond the constituent quark model.
 NRQCD provides an effective factorization formula for the double-charmonium production, where the short-distance
 Wilson coefficients and the LDMEs are well-factorized, since the bottom quark mass is a large
scale compared with the hadron scale $\Lambda_{\mathrm{QCD}}$. But we should also notice that the factorization
to all orders of the strong coupling $\alpha_s$ and all orders of the quark relative velocity $v$ in Upsilon hadronic decays into double charmonia is still needing further investigation.

To understand the data carefully, one should clarify the nature of distinct hadron states. Up until now,
the physics information of the $X(3940)$ and $X(4160)$ is lacking except for their masses and poorly studied
decay modes.
The explanations of the $X(3940)$ and $X(4160)$ as the $3 ^1S_0$ and $4 ^1S_0$ charmonium states, respectively
can explain the experimental upper limit for $\Upsilon(nS) \to J/\psi+X(M)$. The branching
fractions  are around $10^{-7}$ when we treat the $X(3940)$ and $X(4160)$ as a pure charmonium,
and  future measurement with more data by Belle Collaboration will be able to clarify it is or not.

\section*{Acknowledgments}
The author thank Professor Xiangdong Ji and Professor Cheng-Ping Shen for helpful discussions.
This work was supported in part by a key laboratory grant from the Office of Science and Technology,
Shanghai Municipal Government (No. 11DZ2260700),
by Shanghai Natural  Science Foundation  under Grant No.15ZR1423100,
and by the Open Project Program of State Key Laboratory of Theoretical Physics,
Institute of Theoretical Physics, Chinese
Academy of Sciences, China (No.Y5KF111CJ1).

\section*{Appendix}
In our calculation, the package FeynCalc~\cite{Mertig:1990an} is used to generate the Feynman amplitudes, LoopTools~\cite{Hahn:1998yk} is used to calculate the integrals,
FIRE~\cite{Smirnov:2008iw} and Apart~\cite{Feng:2012iq} are  employed to reduce Feynman integrals to Master integrals.
The scalar Passarino-Veltman  Master integrals $B_i$ and $C_i$ are defined in Refs.~\cite{Passarino:1978jh,Hahn:1998yk,Hahn:2000jm}, and
$b_i$ and $c_i$ are related to them with the following identity: $b_i=B_i$, $c_i=C_i/m_b^2$.
\begin{eqnarray}
B_1&=&B_0\left(0,m_b^2,m_b^2\right),\nonumber\\
B_2&=&B_0\left(0,z^2 m_b^2,z^2 m_b^2\right),\nonumber\\
B_3&=&B_0\left(m_b^2,0,0\right),\nonumber\\
B_4&=& B_0\left(m_b^2,m_b^2,m_b^2\right),\nonumber\\
B_5&=& B_0\left(z^2 m_b^2,m_b^2,z^2 m_b^2\right),\nonumber\\
C_1&=&\text{C}_0\left(m_b^2,z^2 m_b^2,z^2 m_b^2,0,0,z^2 m_b^2\right),\nonumber\\
C_2&=&\text{C}_0\left(m_b^2,z^2 m_b^2,z^2 m_b^2,m_b^2,m_b^2,z^2 m_b^2\right).\nonumber
\end{eqnarray}

\end{document}